# Spontaneous Radiofrequency Emission from Electron Spins within *Drosophila*: a novel biological signal.

**Alexandros Gaitanidis[a], Antonello Sotgiu[b] and Luca Turin[c1]**

[a]**Neuroscience Division, BSRC Alexander Fleming, 16672 Vari, Greece**
[b]**Applied Physics, University of l'Aquila, 67100 L'Aquila, Italy**
[c]**Quantum Biosciences, Clore Laboratory, University of Buckingham MK18 1EG UK**

**Abstract**: Using simple radiofrequency (RF) instrumentation, we detect spontaneous RF emission from *Drosophila* immersed in a magnetic field within a waveguide or a shielded RF resonator in the absence of external energy input. Remarkably, the RF emissions are abolished by chloroform anesthesia. Conversely, activation of the nervous system by temperature-sensitive cation channels causes bursts of radiofrequency emission. Both observations suggest that the radiofrequency emissions come from the nervous system. RF frequency and magnetic field dependence are consistent with RF emission being related to electron Zeeman energy. Since the RF emission occurs without external energy input, its energy must come from cell metabolism. We propose that RF emissions are due to *in vivo* spin-polarised cellular electron currents. We suggest that spin-polarised currents relax radiatively to equilibrium at the Larmor frequency and are detected under our experimental conditions (resonator cavity and lock-in detection).

**Introduction**.

**Chirally induced spin polarisation** The work of Naaman and others[1–5] has shown that electrons traversing chiral phases such as proteins DNA, and more recently electron-conducting bacterial fiilaments [6] become spin-polarised in the direction of travel. In aerobic organisms, large electron currents (100 A in a human at rest[7], 1 µA in a single 1-mg fruit fly[8], both under approximately 1.4V) reduce dioxygen to water. These currents flow through chiral proteins, and their spin polarisation by the same mechanism is therefore likely.

 While the surprisingly large extent (as much as 90%) of spin polarisation has been confirmed in a variety of experiments where this variable could be measured directly, the mechanism causing the polarisation remains controversial. The original proposal [1] of large magnetic fields generated by electrons flowing in a tight spiral in the alpha helix backbones seems at present unlikely to be correct., and other mechanisms have been proposed[5,9–14]. It is not yet clear whether spins are being filtered or altered. In the former case, electrons of the wrong spin are unable to flow through the chiral phase and are reflected. In the latter, all spins enter the choral phase and their spins become aligned while transiting in the chiral phase.
Whatever the mechanism, the polarisation cannot survive when electrons exit the chiral phase when electrons relax to thermal equilibrium. This relaxation may in part proceed radiatively, i.e. be accompanied by emission of (microwave) photons. However, for this to happen some anisotropy must be present. If the proteins conducting the electrons are isotropically oriented, the spin-polarised electrons will, from the standpoint of an outside observer, still be pointing equally in all directions and therefore unpolarised. What is observed experimentally in a magnetic field is that electron currents flowing with the field are increased and currents flowing against the field are decreased. This makes it possible to create an overall spatial polarisation. Note that the coupling between spins and magnetic field involved in spin polarisation is small, of the order of 0.1 meV per Tesla. What makes spin polarisation hard to achieve by magnetic field alone is that thermal energy is approximately 25 meV. A magnetic field exceeding 250 T would therefore be required to achieve full polarisation. Conversely, the energies emitted by radiative relaxation to equilibrium will take place in the microwave range, namely at 28 GHz/Tesla for spins with a g-factor of 2 (free electrons or carbon based radicals). Other, larger g-factors are seen in spins attached to atoms other than carbon and these would give emissions at lower frequencies.

---
[1] To whom correspondence should be addressed, luca.turin@buckingham.ac.uk

We surmised that spin polarisation driven by metabolism, i.e. *without external energy input*, could perturb the thermal equilibrium of *in vivo* spin populations and lead to resonant radiofrequency *emission* in a magnetic field. The energies involved in spin polarisation are sufficiently small that they will not appreciably burden the battery supplying the current. Mitochondrial electron currents are driven by an electrochemical gradient of ≈1.4 V, i.e. 1.4 eV per electron, whereas spin polarisation in a magnetic field of, say, 0.1 T involves energies of the order of 10 μeV. RF emission is a well-known consequence of spin disequilibrium, seen in pulse ESR after brief high-intensity RF pulses[15], in chemically induced spin polarization[16,17], and, most similar to CISS, in spintronic devices[18]. It is worth emphasizing that all ESR techniques rely on the *external* supply of energy in the form of RF radiation, whereas our measurements rely on energy *internally* generated by metabolism to generate the emission. It is also important to emphasize that magnetic fields per se do not contribute energy to the system, as can be inferred from the fact that electromagnets can be replaced by permanent magnets, and these of course remain magnetised indefinitely, showing that no energy loss is involved.

For a single fruit fly, an upper bound of the total energy released in the process of relaxation is given by the product of fully polarised electron current and free electron Zeeman energy in the applied magnetic field (115 μeV/Tesla), say 100 pW for a 1 μA current at 1T. Even if only a small fraction, say $10^{-4}$, of this energy, were radiated in response to an easily achievable magnetic field of .1 T, the emitted power would be of the order of a few femtowatts at 3-4 GHz. Our previous electron spin resonance spectroscopy (ESR) studies on electron spin in Drosophila[19,20] suggested fruit flies might be a good organism to look for emission, given their good transparency to radiofrequency even in the X-band, their ease of handling, their sensitivity to anesthetics and the powerful genetics tools available for their study. Commercial continuous-wave (CW) ESR spectrometers are ill-suited to study emission, however, because the incident RF beam can typically be turned down but not off, RF shielding is not optimal, and the cavity is tuned for maximum Q, which is undesirable for emission experiments where sample absorption must be kept low, not maximised. We therefore built two setups specifically designed to measure RF emission in fruit flies. Here, using this new instrumentation, we report the detection of nonthermal RF emitted from a living organism.

**Detection of radiofrequency emissions** A spin-flip transition is a change in magnetic moment and its most efficient coupling to the electromagnetic near field will be magnetic. To detect it therefore requires either a magnetic loop antenna or a resonator arranged in such a way as to achieve a good magnetic coupling in part of its cavity. The advantage of a highly tuned resonator is an increase in sensitivity insofar as the signal builds up in the resonator circuit. Given the small size of the flies we were able to situate the sample within the resonator itself and we used a reentrant-cavity type of resonator [21] operating in the S band (2-4 Ghz) with which a high quality factor together with excellent separation of magnetic and electric fields can be achieved.

The S band is a busy part of the electromagnetic spectrum, used by cordless phones, wi-fi, Bluetooth and microwave ovens. To achieve a good signal-to-noise ratio, in addition to careful shielding (see methods), we used lock-in detection to select the signal of interest[22]. Lock in (a.k.a synchronous) detection is a technique whereby the signal of interest, buried in noise, is modulated at a known frequency $f_r$ and phase by a controlling variable (here magnetic field) , and the modulation signal is fed as a reference to a receiving amplifier which "mixes", i.e. multiplies the noisy input signal and the reference. Multiplying two sinewaves of frequencies $f_1$ and $f_2$ gives two signals: one at $f_1 - f_2$ and one at $f_1 + f_2$ . In our case $f_1 = f_2 = f_r$ so the output signal will contain a DC signal proportional to the input signal oscillating at $f_r$ and a $2f_r$ component which can be filtered out with a low-pass filter with an appropriate cutoff frequency $f_{LP}$ between DC and $2f_r$. In these experiments, the modulation signal $f_r$ is a small sine modulation of the magnetic field at a few kHz. All the frequency components $f_s$ in the input signal which do not match the modulation frequency will appear on the output as $f_s-f_r$ and be filtered out. The time constant $1/f_{LP}$ (for a single pole filter) then acts as the response of the system to an instantaneous change of signal. Overall lock-in amplification is equivalent to a very sharp filter tuned to the frequency $f_r$.

**Methods** The magnetic field in which the flies are immersed consists of two components: a fixed or slowly-varying field provided by a bench electromagnet (DSXG-100, Dexing Magnet) powered by a F-2030 power supply; and a modulation field provided by one or two solenoids (see below), fed from a sine wave generator via a 100W class-D audio amplifier (ST Microelectronics TDA7498) through a series resistor and capacitor to achieve the desired LC resonant frequency of ≈14 kHz. The sine wave signal serves as a reference for the lock-in amplifier. Following continuous-wave (CW) EPR practice, to study magnetic field dependence and improve the signal/noise ratio, we add a modulating magnetic field component to the slowly varying background field and measure the modulated emitted RF using lock-in detection. We are interested in detecting signals which vary in a resonant fashion with the magnetic field. In conventional CW ESR, this is achieved by lock-in detection at the magnetic field modulation frequency to obtain a differential signal proportional to the slope of the underlying absorption curve. Given the uncertainty about the signal phase in our experiments, we used R-Φ (modulus-phase) recording mode and only recorded modulus.

**C-Band (4.5-4.8 GHz) waveguide setup.** We enclosed Drosophila melanogaster wild type (W1118 strain) flies in a 4 mm ID PTFE tube situated at one end of a section of WR229 waveguide. To allow the WR229 waveguide to fit between the poles, the electromagnet pole pieces were ground from the original 30 mm separation to give a 36 mm pole gap and the end flange of the waveguide was trimmed along the short dimension to 34 mm to fit. The waveguide is terminated at the other end, approx 1 metre of waveguide away, by a C-band low-noise block (LNB, New Japan Radio model NJS8488U). The LNB mixes incoming RF at 4.5-4.8 GHz with the signal of a local oscillator at 5.76 Ghz to produce a down-converted signal at 900-1200 MHz with 59-66 dB gain. That signal is fed to a logarithmic power detector (Minicircuits ZX47-60LN-S+) and on to an Ametek 7265 DSP Lock-in amplifier.

The modulating magnetic field is external to the waveguide and is provided by a solenoid made of 0.8 mm diameter copper wire (inductance 530 μH) in series with a 490 nF capacitor. The coil is wound around a ferrite pot core to confine the magnetic field lines and avoid losses in the electromagnet pole pieces. The solenoid-pot core assembly is held on the end of a brass tube clamped in a micropositioner to allow precise three axis-movement. In order to avoid as far as is possible attenuation of the modulation field by eddy currents, the waveguide is terminated by a window consisting of a piece of aluminum foil glued to a 2mm-thick PTFE plate cut to the size of the flange. The ferrite core, while necessary, introduces some additional complications. First, it will concentrate field lines within itself and dilute them elsewhere, so that the transverse DC field in front of the core will be lowered. The field inhomogeneity is relatively unimportant in our case because in experiments using a single fly the size of the specimen (2mm) is still small compared to the field gradient. Second, core inductance will decrease with increasing magnetic field as a result of gradual saturation of the ferrite, and therefore the resonant frequency of the LC circuit made up of the solenoid and series capacitor will fall. This is compensated by lowering the quality factor of the LC resonant circuit by adding 3.3Ω (100W rated) series resistance. Third, the magnetic field produced by the core at a given amplifier gain setting will vary with DC magnetic field. Insofar as peak positions rather than absolute amplitudes were of primary interest, this did not need addressing at this stage.

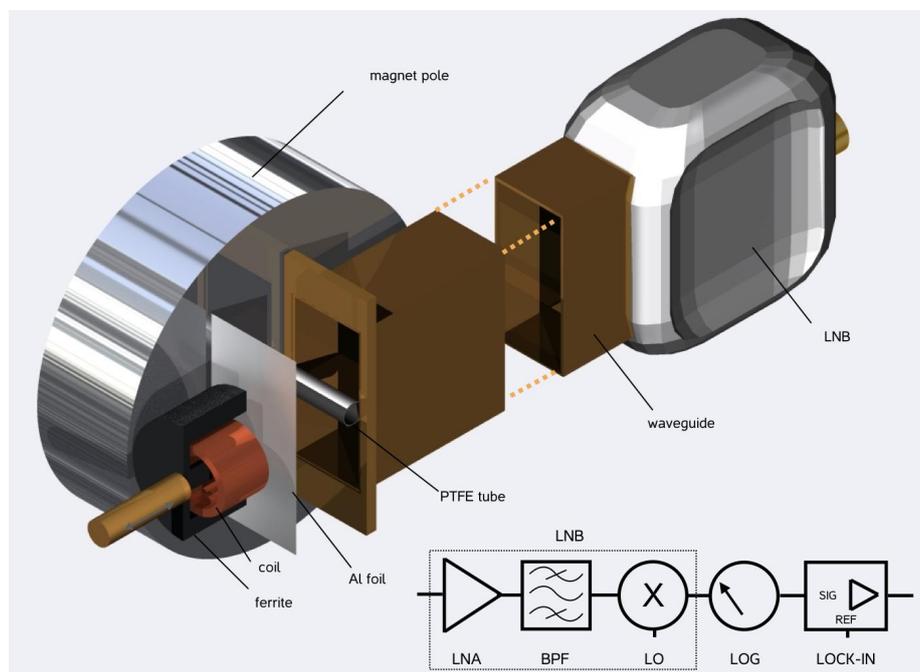

**Figure 1:** Exploded diagram of the waveguide experimental setup. Not shown: the PTFE backing of the aluminum foil, the mirror-image second pole piece of the electromagnet, and a thin sheet of plastic for electrical insulation between the waveguide and pole pieces to avoid current loops. The total WR-229 waveguide length is about 1 m and includes a 90-degree turn. The ferrite pot coil is mounted on a micropositioner for three-axis adjustment. Then signal chain (bottom right) consists of a low-noise amplification stage, a stripline bandpass filter, and a mixer with a local oscillator (LO), all within the LNB, followed by a logarithmic power meter (Minicircuits ZX47-60LN-S+) and an Ametek 7265 DSP lock-in amplifier.

The RMS noise measured on the power detector is approximately 50 mV at an average signal of -1.36 V corresponding to approximately -30 dBm power, with a gain slope of -25 mV/dB. The gain of the setup was calibrated by replacing the endplate with a WR229 flange-to-SMA adapter and feeding it with a fully AM-modulated RF signal (from a Windfreak Technologies SynthHD USB RF generator) attenuated to -120 dBm, to give an approximate power of 1 femtowatt. The main advantages of the waveguide setup are excellent RF quiet provided by the closed waveguide, ease of calibration, and separation between the heat produced in the modulation coil and the fly sample. Its main disadvantages are that the magnetic coupling of the flies to the waveguide is sensitive to distance from the endplate and exact position, as is the coupling of the flies to the modulation. When performing experiments, there is no independent way to make sure that both couplings are good and measurements are therefore hit-and-miss. Another disadvantage is that it is not easy to extract the fly tube, and put it back in the exact same position, or to pass gases through it.

**S-Band (2.6 GHz) resonator setup.** We used a resonator of a re-entrant Φ-shaped design[23] built for these experiments by www.imagtech.it. It is tuned to approximately 2.6 Ghz, and achieves high separation between electrical and magnetic coupling such that the central cavity (I) is uniformly coupled magnetically while being situated at an electric field minimum, one of the side arms (O) being used for coupling ). A PTFE tube containing approximately 10 flies is inserted in the central tube. The coupling loop connected to an SMA snap-on plug in the coupling arm is set to overcoupling in order to maximize energy output from the cavity. The dimensions and dielectric properties of the cavity determine its tuning frequency, which in our case was in the range of 2.6 Ghz, close to frequencies used by cell phones and WiFi .

Our intention was to avail ourselves of cheap accessory electronics in this frequency range, but in retrospect, this design choice turned out to have disadvantages. This is a busy part of the spectrum, and much ambient noise came from WiFi and cellphones. We found it necessary to encase the resonator in an earthed metallic box machined out of aluminum, incorporating holes for modulation current input to the two solenoids in a Helmholtz arrangement, an exit hole for a snap-on SMA signal connector from the coupling loop, and two flanged holes for air inlet and outlet to effect cooling.

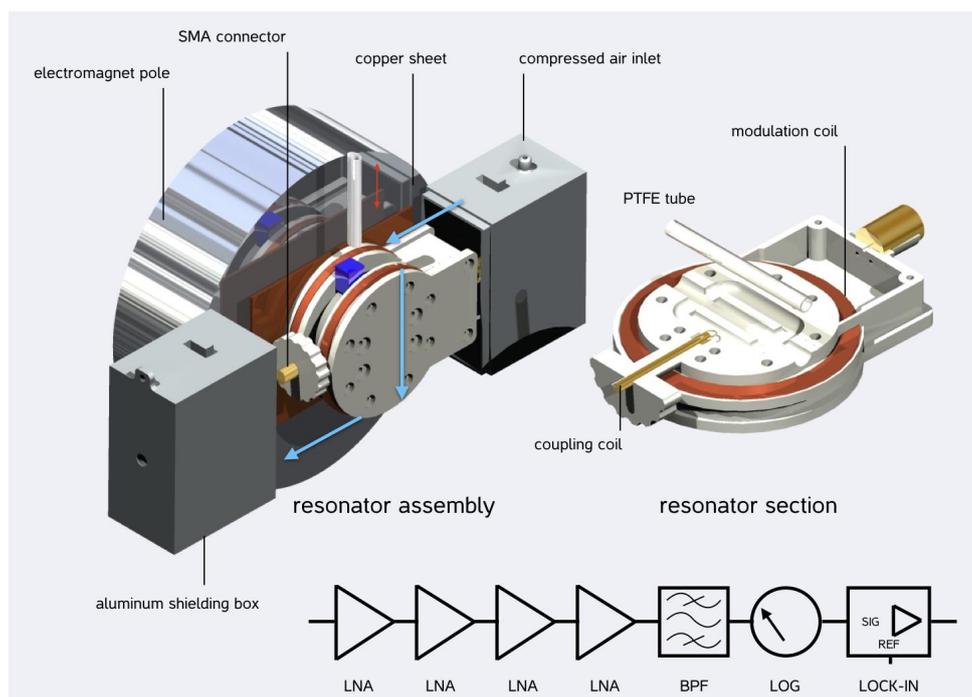

**Figure 2: Left:** Exploded diagram of the resonator experimental setup. Not shown: the mirror-image second pole piece of the electromagnet, and a second copper sheet for cooling. Air enters and exits the shielding box through symmetrically placed inlet and outlet (only inlet is visible). Airflow (blue arrows) within the box is constrained by plastic blocks indicated in dark blue to flow down the resonator tube and along both halves of the box before exiting from the bottom outlet. Right: the resonator sectioned down its middle, showing the shape of the resonant cavity, the position of the adjustable coupling coil in one of the arms and the connector to the moduaton coils at the back. The signal processing chain is shown at bottom right it consists of four low-noise amplifiers (Minicircuits ZX60-P103LN+) in series, a 2.45-2.7 GHz band-pass filter (Microdyne), a logarithmic power meter (Minicircuits ZX47-60LN-S+) and an Ametek 7265 DSP Lock-in amplifier.

An advantage of this setup are that the design of the resonator ensures that the flies are optimally coupled whatever their position inside the resonator. A disadvantage is that the need for holes in the shielding box necessarily allows some RF noise in. Another disadvantage is that the heat from the modulation coils must be dissipated to make sure the flies are kept at a temperature lower than 28-29C. To this end two copper sheets .5 mm in thickness were glued to the side of the resonator and thermally connected to the metal retaining screws with thermal paste. In addition, two small wedges of sponge were positioned in the resonator grooves in such a way as to ensure that air circulated down the central tube, both around and within the PTFE tube containing the flies. To this end, the flies were constrained in the middle two-thirds of the tube by small PTFE plugs triangular in shape to allow air flow without letting flies escape. For exposure to chloroform, the fly tube was removed from the resonator, exposed to chloroform vapor for a few seconds and replaced in the tube.

**Results.**

**General remarks**: In our initial experiments, signals were *not always present* in every batch of live flies. We believe that normal, background activity in the fly nervous system may on occasion be too small to be detected. The flies are in darkness with no external stimuli, and their nervous systems may be minimally active. This is consistent with the fact that experiments in which the nervous system is activated by temperature-sensitive ion channels expressed in the neurons give large, reliable signals in a burst-like pattern. Whenever RF emission is observed in resonator experiments, we believe it to be a genuine signal for the following reasons: **1-** We have now observed the emitted RF behaving similarly in three different setups, the first one using permanent magnets for steady field and lower modulation frequencies[24], as well as the two electromagnet setups described in this report. **2-** Long-term control recordings of the resonator lasting up to 48 hours (figure 6) show no events resembling the signal. **3-** The dependence of the signal on magnetic field exhibits resonant peaks (figures 3,4 and 5). It is hard to see what artifact could be resonant with a slowly-varying magnetic field **4-** The signal is only detected when modulation is on (figure 7A) **5-** The signal only occurs when the flies are alive and disappears gradually as they die (figure 7B) **6-** The signal disappears in flies exposed to chloroform vapor (figures 4 and 5), **7-** Transgenic flies expressing temperature sensitive cation pan-neuronally but not in other tissues give large-burst-like RF signals (Figures 8).

**Waveguide experiments at 4.5-4.8 GHz.** Figure 3 shows the best trace obtained to date with the waveguide setup. The exceptionally quiet RF environment, high gain and low noise figure of the LNB-waveguide arrangement afford an excellent signal/noise ratio. The trace obtained a few minutes earlier with an empty tube (black and inset) shows gaussian noise filtered by the output time constant of the lock-in amplifier (1 second). The red trace with live flies shows several clear resonant peaks. Despite the unquestioned advantages in radio quiet and signal/noise ratio, we felt that the lack of reliable relative positioning of the solenoid, end-plate and flies added an unnecessary difficulty to the experiments, and we focused on the resonator setup.

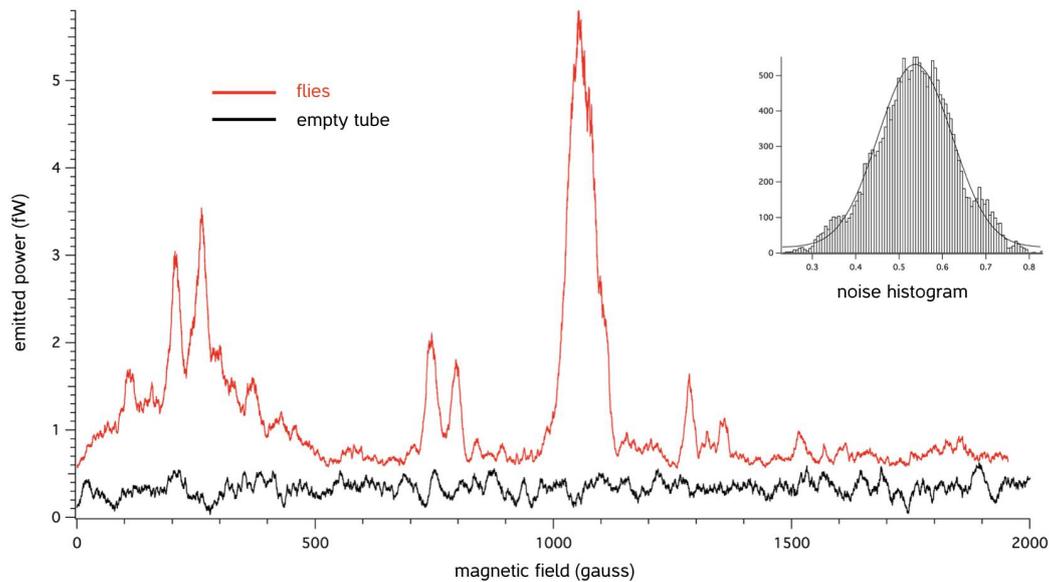

**Figure 3:** Radiofrequency emission from ≈ 10 fruit flies contained in a Teflon tube positioned at the end of a WR229 waveguide, detected at 4.5-4.8 GHz with a room temperature C-band satellite receiver. The value of the steady magnetic field is given in the abscissa, emitted RF power in the ordinate. The signal is detected using a logarithmic power meter and a lock-in amplifier, with a modulation frequency of 16 kHz, a width of ≈90 gauss and a response time constant of 1 second (6dB per octave). The steady magnetic field is ramped at approximately 10 gauss/second. The trace is exponentiated to restore linearity. It clearly shows several resonant features at 200, 270, 770, 800, 1050 and 1300 gauss. The red trace shows the signal when live flies are present in the tube. The black trace is taken a few minutes earlier with no flies in the tube and has the properties of gaussian noise (see inset histogram of black trace). The empty tube trace has been shifted downwards in the main graph by 0.2 fW for clarity.

With the exception of the heat activation experiments, flies for all experiments were wild type flies (Canton-S) and raised at 25C. For the neuronal activation experiments with TRPA1, flies were raised at 18C. All flies were tested 3–10 days after emergence. For neuronal activation experiments, expression of UAS-dTRPA1 was driven to the neurons using the paneuronal driver elav-Gal4.

### Resonator experiments at 2.6 Ghz

**Slow recordings.** The resonator setup has the advantage of a high degree of magnetic coupling to the fly at the expense of greater RF interference because it does not in itself provide a Faraday cage. Our initial design criteria planned for interfacing with generators and circulators, to eventually build a full ESR setup. 2.6 GHz was chosen to ensure that such components were available and affordable, insofar as they are mass-produced for cellular and WiFi use. This turned out to create its own difficulties, because that RF band is busy, and the low-noise measurements demanded by our experiments required close attention to additional shielding. We therefore designed a tight-fitting Faraday "box" in two parts with a lip overlap. Bored holes at each end to allow connection of the modulation and the amplifier chain. This greatly reduced noise, and in practice allowed measurements to be made with a lock-in output filter time constant of 10 seconds, 6dB/octave. Modulation frequency was set to 13.9 kHz, the resonant frequency of the 376 μH solenoid coils in series with 455 nF. Resonance was determined by using an external pickup coil made of a few turns of copper wire and setting frequency manually to the (rather shallow) maximum magnetic field value. The modulation coils give 20 gauss/A, and large values of modulation were found to be necessary to detect the signal, a fact which may be important in interpreting the results (see discussion). Conventional Class D and class AB audio amplifiers struggled with these frequency and current levels and sometimes gave parasitic oscillations leading to excess heating in the solenoid. We eventually settled on an inexpensive class D amplifier capable of a maximum output of 7A at 14 kHz without overheating or damage. In the experiments described below the maximum modulation of 140 gauss was used.

An accidental discovery increased the chance of observing a signal: while we were dealing with airflow within the box to keep the temperature within limits tolerable by the flies, we noticed that a brief period of nonlethal heat stress around 30C caused the flies to emit signal more reliably. We did this deliberately in a number of experiments. Figure 4 illustrates one such experiment. The black trace (shifted downwards by .1 units for clarity) was taken with an empty tube. The first trace with flies, slightly above normal temperature is shown in blue and a small signal is visible around 750 and 850 gauss. The second trace 181 minutes later is shown in red and a large signal is clearly visible around 800 gauss. Note that practically the entire red trace is above background, and that smaller signals can be seen both below and above the main peak. After exposure to chloroform, the signal returns to baseline (green trace, also shifted downwards by .1 units for clarity).

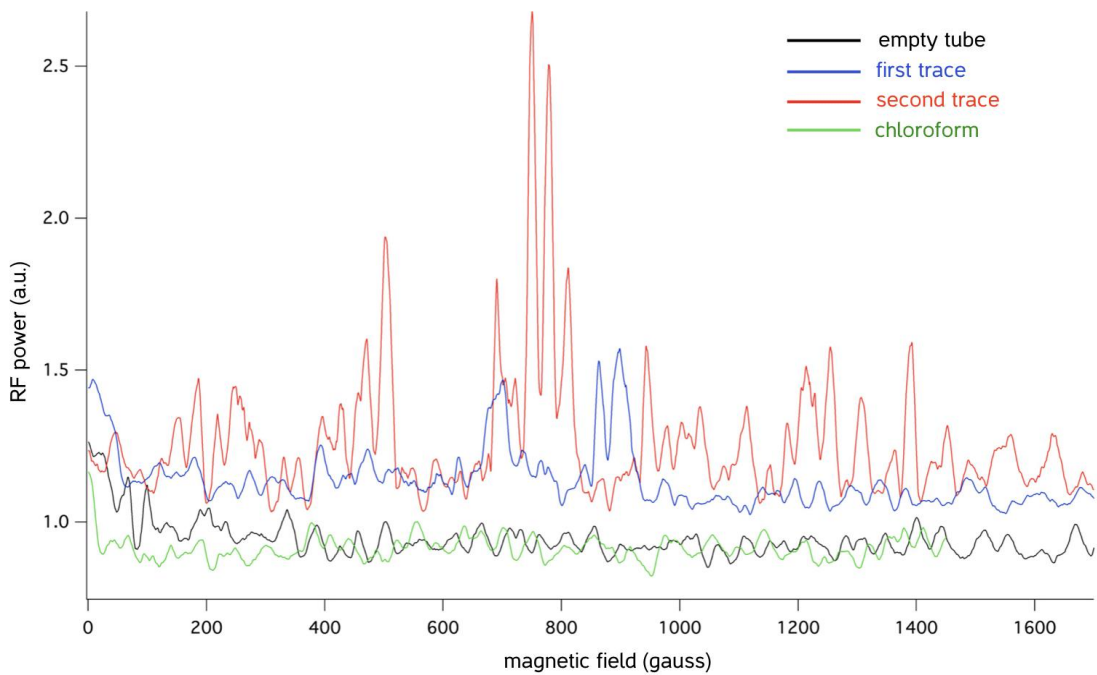

**Figure 4:** Four traces from a heat-stress experiment in the 2.6 GHz resonator. The bottom two traces have been shifted downwards for clarity. See text for details. The sweep from 0 to 2000 gauss took 30 minutes, and the lock-in output time constant was 10 seconds, 6dB/octave. The ≈18000 point traces were smoothed with a 100-point box filter after acquisition for noise reduction. The ordinate is in arbitrary units because calibration of the resonator setup is difficult insofar as the degree of coupling between cavity and pick-up coil is unknown. See figure 3 for a better measurement of likely power output.

A similar but larger signal is shown in figure 5. Conditions are the same as in figure 4, i.e. 30-minute sweep and 10-second output time constant. The fliers were subjected to a brief heat stress of 5 minutes. Most of the signal occurs between 650 and 1200 gauss. Fluctuations in the peak are larger than before and after, suggesting that the peak may have some substructure unresolved under the current measuring conditions. A brief exposure to chloroform 35 minutes later abolishes the signal (black trace).

RF emission from *Drosophila*

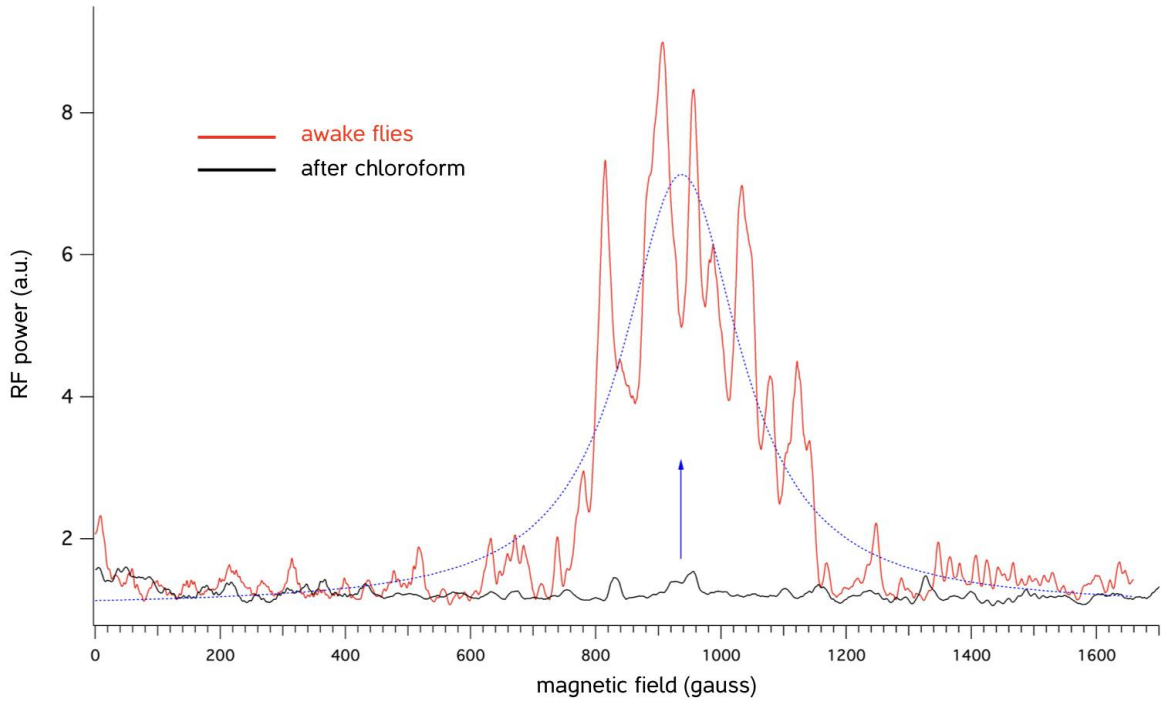

**Figure 5:** The best trace obtained to date in the 2.6 GHz resonator. See text for details. The sweep from 0 to 2000 gauss took 30 minutes, and the lock-in output time constant was 10 seconds, 6dB/octave. The ≈18000 point traces were smoothed with a 100-point box filter after acquisition for noise reduction. The ordinate is in arbitrary units because calibration of the resonator setup is difficult insofar as the degree of coupling between cavity and pick-up coil is unknown. The dotted blue line is a Lorentzian fit with a midpoint of 930 gauss (arrow).

Naturally, the question arises of whether the signals shown in figures 4 and 5 are merely random glitches of exceptional magnitude that happen to take place during a sweep. The radiofrequency environment is very variable, and occasional large pulses are seen in the log detector trace, so this possibility cannot be excluded *a priori*. To test for this we made long-term recordings of the signal under the same conditions as figures 4 and 5 with an empty tube and modulation on over a period of 17.8 hours. The complete trace is shown in Figure 6A. No transients above noise are visible. We are therefore confident that the signal in figure 5 is not due to a random external perturbation.

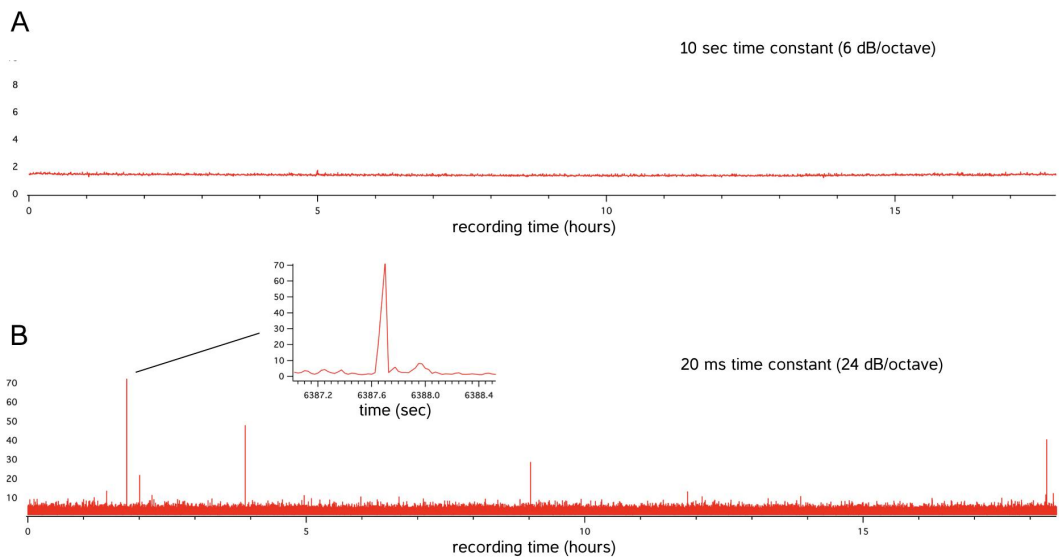

**Figure 6: A:** Continuous 17.8 hours recording of lock.in signal output with empty resonator under the same conditions and at the same scale as the experiment depicted in **figure 5** (red trace). The trace shows no activity at all, and the noise level is identical to that observed in chloroform in figure 5 (black trace). **B:** Continuous 18.4 hours recording of lock-in signal output with empty resonator under the experimental conditions used for **figure 7A**, with modulation on. 8 glitches above noise are visible in the trace, i.e. approximately one every 140 minutes. All of them behave as the largest one does (inset), i.e. last.≈100ms. The rest of the trace is baseline noise. Over 18.4 hours, the signal integrates to $2.10^4$, as compared to $3.68\ 10^5$ in 3 minutes for the experiment depicted in 7A, red trace.

**Fast recordings.** A further improvement in background noise was obtained by wrapping the resonator Faraday box in a bag made of two layers of conductive RF fleece (Aaronia 100-dB X-dream) for additional shielding. The fleece is sufficiently porous that the air outlet is unimpeded. Cable and tube exits from the fleece bag were secured with ring ties. Figure 6B shows the noise trace obtained over 18.4 hours with this method using a much shorter (20ms) lock in time constant.

This shortening of the lock-in output time-constant from 10 seconds to 20 ms (24 dB/octave filter) allowed us to measure the signal with higher bandwidth. Previous experiments [24] using permanent magnets for the steady magnetic field had provided hints that the signal came in short pulses lasting a fraction of a second. Such pulses would be smoothed into a noisy continuous trace under the recording conditions of figures 3, 4 and 5, but could possibly be resolved with a shorter time constant. Figure 7A shows two successive 3-minute magnetic field sweeps taken on the same set of flies, first without (black), and then with modulation (red trace). The black trace is indistinguishable from gaussian noise. A striking increase in the number of short RF pulses is seen when the modulation is turned on. Compare with the long-term trace taken with modulation but without flies shown in figure 6B under the same measurement conditions.

In order to assess the pulsed signal more quantitatively, the trace was integrated after setting the zero to the median of the trace obtained without modulation, such that gaussian noise alone would integrate to zero. The integrated traces are shown in figure 7B. The red and black traces correspond to the traces in figure 7A, the orange trace is taken 8 minutes later with the modulation on, the yellow one 5 minutes later. The flies are thermally stressed, and the signal is decreasing. The next traces (orange and yellow) give a decreasing signal, and finally one indistinguishable from baseline (not shown). Upon opening the resonator, the flies are found to have died. The power spectra of the signals with and without modulation show an approximately 100-fold difference in signal power at low frequencies, gradually decreasing as frequency rises, but remaining detectable all the way to the Nyquist frequency of these measurements, 2.5 kHz.

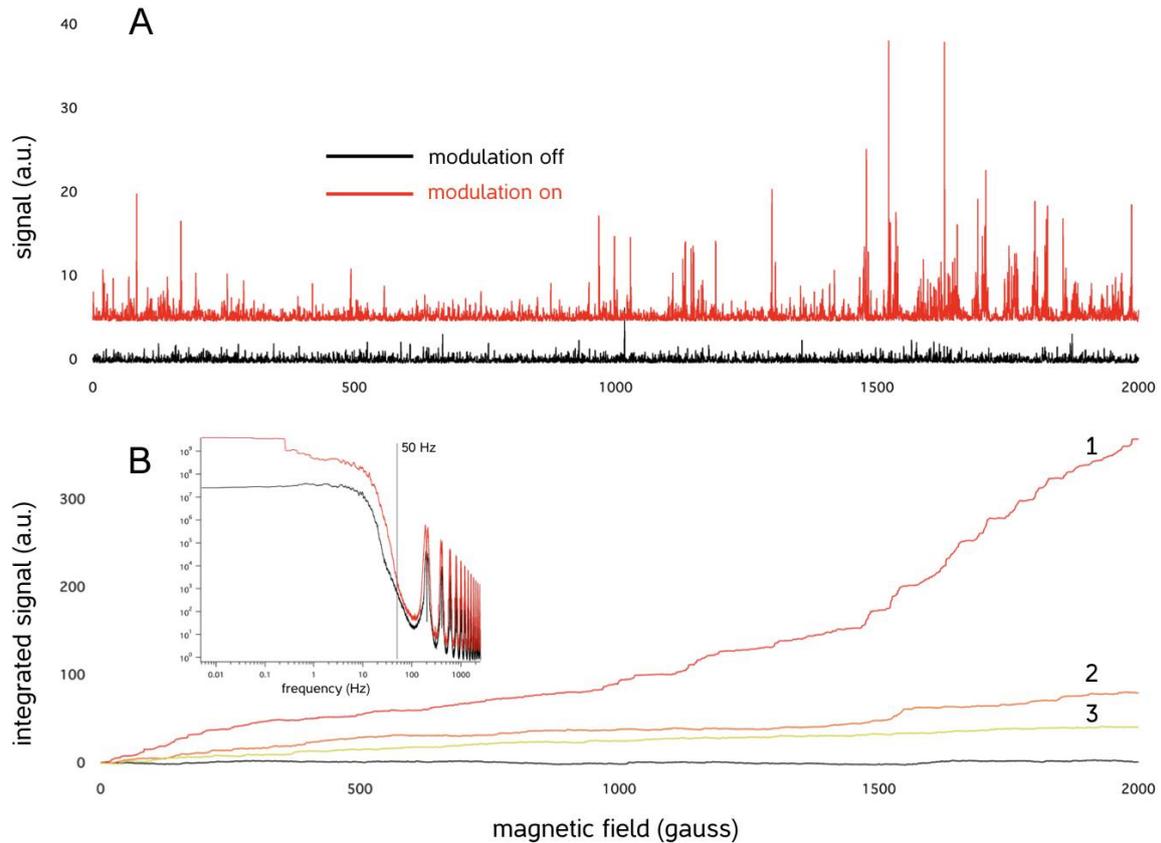

**Figure 7: A**: Two 3-minute magnetic field sweeps recorded with a 20 ms lock-in output time constant. In black, first sweep without modulation. In red, immediately following, with 140 gauss modulation (and therefore heating) switched on. Sampling frequency 5 kHz, 16 bits resolution. A marked increase in pulse-like signals is seen when the modulation signal is on. The red trace has been shifted up by 5 units for clarity. **B:** the traces in A integrated from T=0 to 180 seconds. Trace 2 (orange) is taken 8 minutes after the red one and shows the signal decreasing in amplitude, trace 3 (yellow) 5 minutes later.. **Inset:** power spectra of the two traces in A. The power spectrum (magnitude squared) is dominated by the output filter time constant, but the signal power with modulation on is two orders of magnitude larger at low frequencies and remains above the no-modulation spectrum at all frequencies up to and beyond the nominal cutoff frequency of 50 Hz, including in the stopband ripples above 100 Hz. This suggests that the signal added by modulation has a wider bandwidth than current measurements allow.

### Radiofrequency emission during activation of the nervous system by temperature-sensitive cation channels

In order to test whether the observed radiofrequency emissions originated from the flies' nervous system, we expressed temperature-sensitive cation channels pan-neuronally using a neuron specific driver. The transient receptor potential cation channel, subfamily A, member 1 (TRPA1), is a membrane ion channel. dTRPA1, the *Drosophila* ortholog, is a temperature- and voltage-gated cation channel [25,26] that regulates *Drosophila* thermotactic behavior [27,28]. Previous work has shown that neurons expressing dTRPA1 begin firing action potentials when temperature rises above 25C [25](Hamada et al., 2008). By ectopically expressing dTRPA1 using the GAL4-UAS system, and then delivering modest heat pulses, sets of neurons can be remotely activated in freely behaving animals [29]. It has been shown that activating all neurons with dTRPA1 causes tetanic paralysis in larvae, while in adults it led to paralysis in males and continuous uncoordinated leg and wing movements in females [30].

The results are shown in figure tktk below. In the first experiment (panels A and B), the teflon tube containing the flies was inserted into the resonator after it had been cooled to 4C. The thermal inertia of the resonator is such that when the current in the modulation coils was switched on, the resonator warmed at approximately 2C per minute.

The temperature is measured at the beginning and the end of the experiment, and linearly interpolated between those two values. Panel A shows a continuous RF trace measured at a constant value of magnetic field (580 gauss) at which activity had been seen during a previous field scan. The trace is quiet until just before 8 minutes, corresponding to approximately 20C when a large spike is seen. The signal gradually grows thereafter, with some more and less active periods, until at 12 minutes it begins to exhibit large, continuous spiking activity. The experiment was terminated after 14 minutes, the temperature inside the resonator was 31C and the flies were alive.

Panel B shows the short-time Fourier transform (spectrogram) of the signal shown in panel A. The output time-constant of the lock-in was 10 ms (100Hz) and the frequency response of the system was therefore conservatively estimated to be 50Hz. A small amount of activity around 28 Hz is visible in the first few minutes before the large spike. It is worth noting that CNS activity at a frequency close to 30 Hz has been described as a correlate of attention in flies[31] The gradual increase in activity after the spike as the temperature rises manifests itself in an increase in power around the previously existing ≈28 Hz signal, with additional bands appearing at approximately 3, 20 and 32 Hz. In addition, broadband noise is visible all the way up to the frequency limit of 50 Hz in the minutes before the experiment ends. This suggests that greater overall bandwidth of the system may reveal higher-frequency fluctuations in the RF signal.

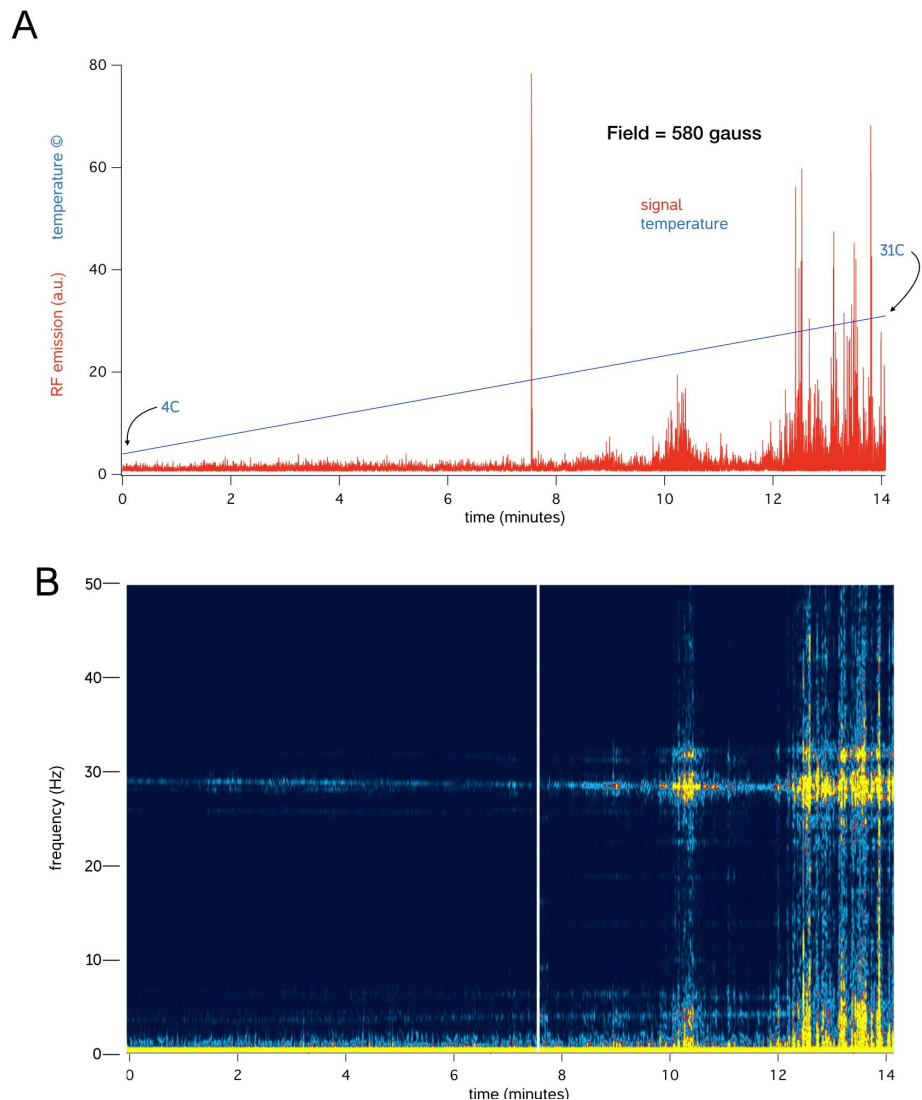

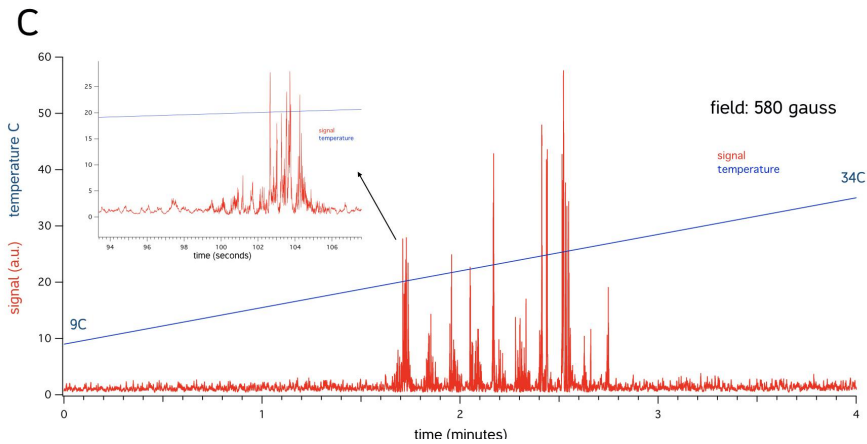

**Figure 8:** A: Fourteen minutes of continuous recording at the fixed magnetic field value of 580 gauss. The resonator was equilibrated in a refrigerator at 4C overnight, and the flies were inserted into it just before the beginning of the trace, with a resonator temperature of 4 C. The modulation was then switched on, and RF emission began to increase after nearly 8 minutes, rising gradually. At 14 minutes the modulation was stopped, the flies were found to be alive and undergoing convulsions, from which they subsequently recovered when returned to room temperature. **B:** The data in panel A analysed by short-time Fourier transform (spectrogram). **C:** A trace from a different experiment also conducted at 580 gauss in which the resonator temperature started at 9C. The flies gave bursts of RF ending at approximately 2.8 minutes. After a further 1.2 minutes the experiment was stopped, the flies were found to be dead and resonator temperature 34C.

Panel C shows a different experiment, in which resonator temperature was 9C at the beginning of the experiment, and 34C four minutes later. The trace was initially quiet, then sudden rhythmic bursting activity began at approximately 1.7 minutes, corresponding to approcimately 20C and ended a minute later. After another minute the resonator was opened, the temperature was 34C and the flies were dead. The inset in the figure shows the first burst with a faster timebase. The seemingly random, fast fluctuations within the burst are clearly visible.

**Discussion.** Many experimental aspects of the signals we report require further study. First, the factor governing the overall presence or absence of signal in control flies remains to be determined. Our experiments with TrpA flies suggest that the resting signals from quiescent flies may be too small to measure. It would be interesting to know whether visual or olfactory stimuli causing CNS activity intermediate in intensity between none and full convulsions will give a graded response. Second, the relationship of signal amplitude to modulation amplitude is unclear at the present time. It is our impression that the signal depends nonlinearly on modulation amplitude and that a threshold may exist for RF generation. If indeed this is spontaneous emission, its rate should go up as the third power of magnetic field[32], and the derivative signal we measure should increase with the square of the modulation field. Third, the origin of the RF pulses, while likely to be in the nervous system, cannot be precisely determined at the moment. The TrpA experiments activate muscles as well as neurons, and therefore do not pinpoint the origin of the signal with complete certainty. We recall that the tube contains several flies. Are they all emitting? Single-fly experiments will be needed to address this question. It should also be recalled that we are measuring signals from a few flies weighing a milligram each. Signals from larger organisms, if they scale linearly with mass, should be much larger and should justify the effort to scale up the experiments to, say, a mouse-sized device or larger.

Radiofrequency emissions from living systems have been previously reported [33–35]. For a comprehensive review of Soviet work see [36]. Magnetic field was not used as a controlling variable in any of these studies. The lack of recent interest is no doubt in part due to the poorly controlled experimental conditions and emphasis on biological radiocommunication of many of these early reports. Emission of radiofrequency does not in any way guarantee its reception, and the present study makes no claims as to the use of RF as a means of communication. We are instead interested in the mechanism by which the signals are generated, and in what they tell us about cellular processes. Radiofrequency emissions from chemical reactions are well understood, but they arise from polarisation of nuclear spins [17,37] and therefore occur at NMR frequencies, one or two orders of magnitude lower.

Our motivation for measuring RF emission in this frequency range and in a magnetic field was the finding, proposed by Naaman and colleagues [1] that electron currents flowing through chiral media become spin polarised in the direction of travel. In vitro experiments have confirmed this in a variety of systems, but the mechanism (chirally induced spin-selectivity or CISS) by which it occurs is still debated[10]. Regardless of theory, if experimentally confirmed *in vivo* by RF emission, the generation of spin-polarised electron currents would bring biological electronics into the realm of spintronics. This would have consequences for biological electrochemistry[38]). It would also affect the interpretation of biological ESR spectra, since absorption of microwaves by unpaired electrons crucially depends on their thermal equilibrium distribution[22].

While our results make a case for the existence of emitted RF, its relationship to CISS remains unclear. On the one hand, a clear dependence on magnetic field is seen, which, together with the RF frequency, is a strong indication that a magnetic species out of thermal equilibrium, with a gyromagnetic ratio close to that of the electron, is responsible for the RF emission. On the other hand, at least in the simplest interpretation of the phenomenon as the reverse of absorption, emission does not occur solely in well-defined regions of the background magnetic field. Some of our resonator data (see figure 5) is consistent with emissions from a species with a Landé g-factor close to 2 ($\approx$ 930 gauss at 2.6 GHz), but there also appear to be emissions at lower and higher field values. The main peak in figure 3, for example, even accounting for magnetic field distortion by the ferrite core, does not match the expected value of $\approx$1600 gauss expected for free electrons at 4.5 GHz. Lower values may be due to higher g-factors, commonly seen in biology, for example when metal ions are involved. Higher magnetic field values than those corresponding to g=2 require a different explanation.

A striking feature of the results reported here is that the RF power emitted by the fruit flies is not the result of energy input into the system: it is instead powered by the flies' own metabolism. If CISS is responsible, the magnetic field in which the flies are immersed merely diverts a small part of the electrochemical force driving polarised electron currents into RF emission. Assuming a total oxidative metabolism energy of 1.4 eV, emission at 4GHz would represent an energy loss to RF photons of 15 µeV, i.e. approximately 10 parts per million. The RF emission we observe can be viewed as a passive, non-invasive measurement of the underlying electron currents, whose functional importance remains to be determined. Its ultimate interest and usefulness will depend on identifying the currents' origin and function in the living organism.

The fact that the RF emission varies rapidly is also remarkable. Some of the variability may be due to changes in the sharpness of the cavity resonance due to movements of the flies, which would have the effect of modulating a constant RF output. However, this does not seem likely to cause the short, transient pulses up from baseline and back that are observed. Most electron currents in living systems flow through mitochondria, and some or all of the RF emission may be thus related to mitochondrial metabolism. A constant RF emission would be expected for example from a steady mitochondrial current flowing to oxygen. A rapidly varying current indicates instead that something more complex is going on. This could conceivably be due to mitochondrial electron currents switching on and off. However, the fact that the signals disappear in chloroform, whereas respiration is only mildly inhibited[20], suggests instead a direct connection with the central nervous system. Note that these two possibilities are not mutually exclusive: there is much evidence connecting the still mysterious action of general anesthetics to mitochondrial function[39–42]. Our previous work on olfaction has established a connection between electron currents, odorant sensing and G-protein activation more generally[43–47], while that on general anesthesia has shown a correlation between spin content of Drosophila (as measured by CW ESR) and general anesthetic action[48,49].

**Contributions.** AG and LT designed and performed the experiments. AS designed the resonator and its amplification chain. LT wrote the paper, AG and AS reviewed the paper.

**Acknowledgments.** We thank Makis Skoulakis, Andrew Horsfield, Ron Naaman and Yossi Paltiel for discussions and encouragement. We are very grateful to Peter Hore and Aharon Blank for advice on theory and experimental method. This project was funded by DARPA Biological Technologies Office (grant N65236-18-1-1000 to LT) and we thank them for their generous support. LT thanks the Stavros Niarchos Foundation for support.